\documentclass{aastex62}
\usepackage[utf8]{inputenc}
\usepackage{amsfonts}
\usepackage{amsthm}
\usepackage{mathrsfs}
\usepackage{bm}
\usepackage{amssymb}
\usepackage{amsmath}
\usepackage{dsfont}
\usepackage{float}
\usepackage{wasysym}
\usepackage{booktabs}
\usepackage{savesym}
\usepackage{mathtools}
\usepackage{amsmath}
\usepackage{dsfont}

\savesymbol{tablenum}
\usepackage[separate-uncertainty = true,multi-part-units=single]{siunitx}
\restoresymbol{SIX}{tablenum}

\DeclareSIUnit\year{a}

\DeclareMathOperator{\sign}{sign}
\received{}

\submitjournal{ApJ}

\shorttitle{Sharp gap edges in dense planetary rings}
\shortauthors{Gr{\"a}tz et al}

\begin{document}

\title{Sharp gap edges in dense planetary rings: An axisymmetric diffusion model}

\correspondingauthor{Fabio Gr{\"a}tz}
\email{fgraetz@uni-potsdam.de}

\author[0000-0002-0947-6377]{Fabio Gr{\"a}tz}
\affil{Institute of Physics and Astronomy,\\
  University of Potsdam,\\
  Karl-Liebknecht-Str. 24-25,\\
  14476 Golm, Germany}

\author{Martin Seiß}
\affil{Institute of Physics and Astronomy,\\
  University of Potsdam,\\
  Karl-Liebknecht-Str. 24-25,\\
  14476 Golm, Germany}

\author{J{\"u}rgen Schmidt}
\affiliation{Astronomy Research Unit,\\
  University of Oulu,\\
  FI-90014 Oulu, Finland}

\author{Joshua Colwell}
\affiliation{Department of Physics,\\
  Physical Sciences Building 434,\\
  4111 Libra Drive,\\
  University of Central Florida,\\
  Orlando FL 32816-2385}

\author{Frank Spahn}
\affil{Institute of Physics and Astronomy,\\
  University of Potsdam,\\
  Karl-Liebknecht-Str. 24-25,\\
  14476 Golm, Germany}



\begin{abstract}
One of the most intriguing facets of Saturn's rings are the sharp edges of gaps in the rings where the surface density abruptly drops to zero. This is despite of the fact that the range over which a moon transfers angular momentum onto the ring material is much larger. Recent UVIS-scans of the edges of the Encke and Keeler gap show that this drop occurs over a range approximately equal to the rings' thickness. \cite{BGT82,BGT89} show that this striking feature is likely related to the local reversal of the usually outward directed viscous transport of angular momentum in strongly perturbed regions. In this article we revise the \citet{BGT89} model using a granular flow model to define the shear and bulk viscosities, $\nu$ and $\zeta$, and incorporate the angular momentum flux reversal effect into the axisymmetric diffusion model we developed for gaps in dense planetary rings \citep{graetz2018}. Finally, we apply our model to the Encke and Keeler division in order to estimate the shear and bulk viscosities in the vicinity of both gaps.
\end{abstract}


\section{Introduction}

Disks are a common structure in the universe that appear on a vast range of different size scales: galaxies, active galactic nuclei, protoplanetary disks, exoplanetary systems, and planetary rings. Recent works have shown that bodies embedded in protoplanetary disks or planetary rings create S-shaped density modulations called propellers if their mass is smaller than a certain threshold or alternatively cause a gap around the entire circumference of the disk if the embedded bodies' mass exceeds this critical mass \citep{sremvcevic2000density,sremvcevic2002density}. Two counteracting physical processes govern the dynamics and determine what structure is created: the gravitational disturber exerts a torque on nearby disk particles, scattering them away from itself on both sides, thus depleting the disk's density and forming a gap. Diffusive spreading of the disk material due to collisions counteracts the gravitational scattering and has the tendency to fill the gap.

In its vicinity, the embedded moonlet creates an additional wake in the ring material. With their \emph{streamline formalism} \citet{BGT89} showed  that the viscous transport of angular momentum locally reverses in this region if the ring is perturbed strongly enough by the moonlet wakes (flux reversal) and a sharp gap edge can form.

\begin{figure}[h]
  \centering
  \figurenum{1}
  \epsscale{1}
  \includegraphics[width=18.75cm]{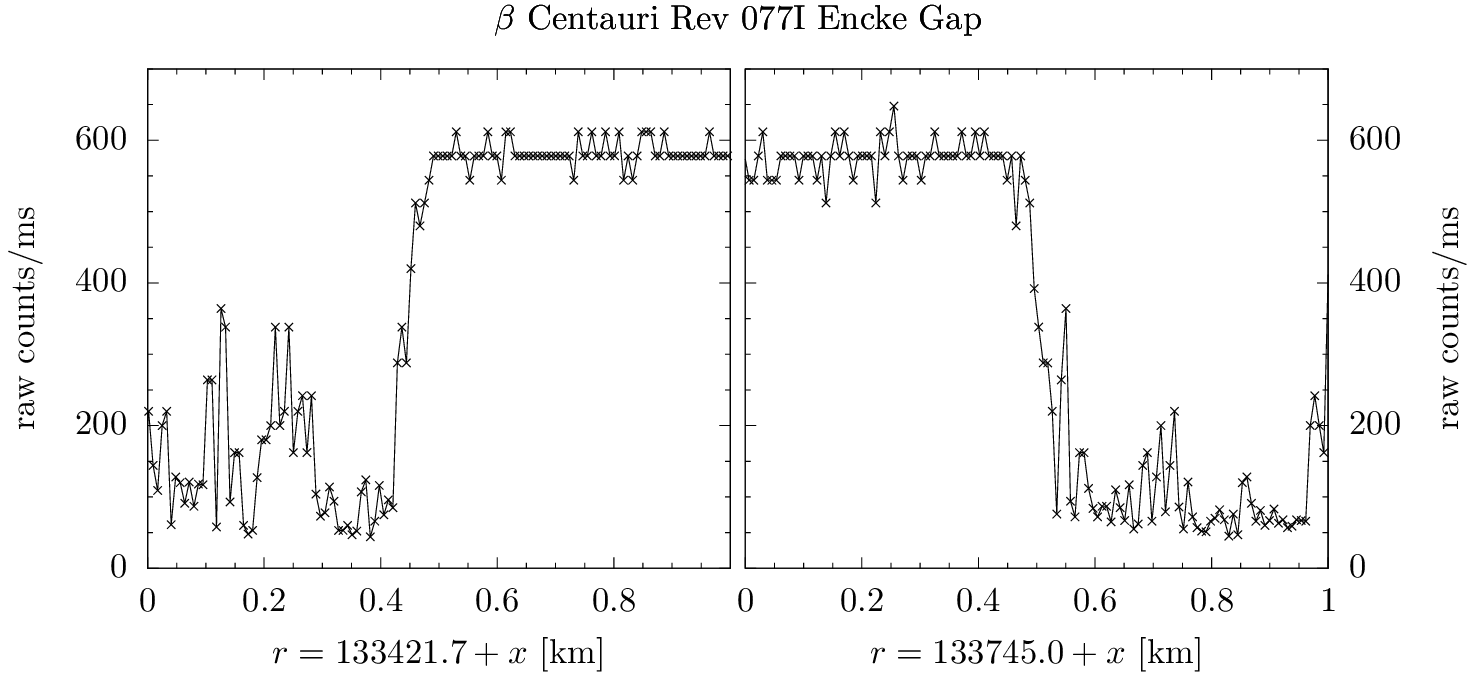}
  \includegraphics[width=18.75cm]{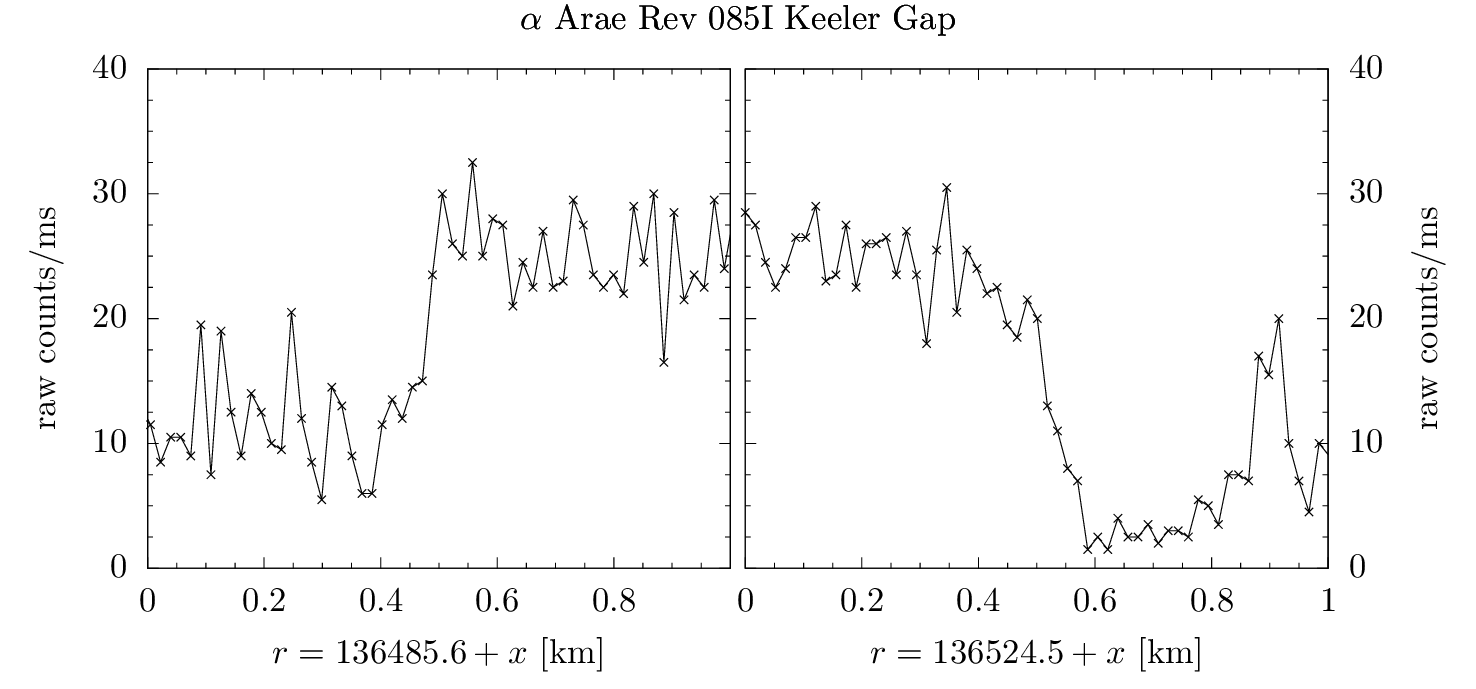}
  \caption{Raw photon counts per millisecond at the inner and outer gap edges of the Encke (top) and Keeler gap (bottom) in the Beta Centauri Rev 077 UVIS occultation (ingress) and the $\alpha$ Arae Rev 085 UVIS occultation (ingress), respectively. The plots show that the ring fluctuates in optical depth on a short spatial scale, which is caused by perturbations from the gap moon as well as resonances with Pandora and Prometheus. Then, at the gap edges, the measured signal increases over a range of less than $\SI{100}{\meter}$ to the photon count rate of the unocculted star. The scans have a radial resolution of around $\SI{8}{\meter}$ (Encke gap), and around $\SI{17}{\meter}$ (Keeler gap), which bespeaks the incredible sharpness of the gap edges. }
  \label{fig:uvis}

\end{figure}

Stellar occultations carried out by the UVIS instrument on board the Cassini spacecraft revealed that the optical depth abruptly drops to zero over a range of less than $\SI{100}{\meter}$ at the gap edges of the Encke and Keeler gaps (see figure \ref{fig:uvis}). This is a distance scale much shorter than that over which the satellite transfers angular momentum onto the ring \citep{BGT82}.

The streamline formalism is a hydrodynamic treatment of ring systems that was developed by Borderies, Goldreich, and Tremaine during the 1980s \citep{BGT82, BGT83a, BGT83b, BGT85, BGT_86, BGT89}. It combines microphysical processes using the Boltzmann equation or semi-heuristic descriptions and the large-scale ring dynamics with perturbation methods from celestial mechanics. The formalism was recently systematically reviewed by \citet{Longaretti1, Longaretti2} explaining that the formalism's most powerful aspect is its ability to deal with a large range of different time scales beginning with the orbital one (usually a few hours), which characterizes the relaxation time of the stress tensor, over intermediate time scales (from one year to tens of years) associated with the synodic period of embedded satellites to the large time scale relating to the viscous component of the stress tensor (hundreds or thousands of years) over which mass and angular momentum are redistributed in the ring.

\citeauthor{BGT82} distinguish between shepherding of a sharp gap edge by an isolated resonance with an outer satellite \citep{BGT82}, which was later analyzed in detail by \citet{hahn2009} at the example of the B ring edge, and formation of a sharp gap edge by overlapping resonances caused by an embedded perturber, which was analyzed in detail by \citet{BGT89} at the example of the Encke gap. The responsible flux reversal was later observed in \emph{N}-body simulations of perturbed rings performed by \citet{HANNINEN1992} and (in a much higher resolution) by \citet{lewis_stewart_2000}: \citeauthor{lewis_stewart_2000} closely analyzed the dynamics of the strongly perturbed region around the Encke gap using local simulations with properly sized particles. They simulated the creation of dense edges, observed angular momentum luminosity reversal, found that the analytic treatments of \citeauthor{BGT82} appear to be generally accurate while discovering additional complexities such as vertical splashing and boundary layer structures that are not encompassed in analytic fluid models of this system.

Gap formation was studied as well by \citet{1981Natur_292_707L} and by \citet{monte_carlo_petit_henon_1988} using a Monte Carlo method and \citet{SHOWALTER1986} examined the wakes created by Pan using images taken by Voyager. \citet{Spahn1989124} used a probabilistic Markov-chain model to describe the scattering process of a perturber embedded in a dense planetary ring. This model was later employed by \citet{sremvcevic2000density} and \citet{sremvcevic2002density} to predict and describe propeller structures that are formed by two counteracting physical processes: \citeauthor{sremvcevic2000density} used probabilistic transitions between radial positions to describe the scattering process in the scattering region and a diffusion equation to model the rings counteracting viscous diffusion back into the created gaps along the azimuth. Propeller structures were then discovered by the Cassini spacecraft (\citet{Tiscareno:2006aa,Tiscareno2008, Tiscareno2010} and \citet{srem_2007}). The ring moons Pan and Daphnis were found in the Encke and Keeler division, where their gravity is strong enough to sweep free a complete gap \citep{Showalter:1991,daphins_entdeckung}.

\citet{stewart91} and \citet{Seiss2010} theoretically analyze collisionless edge dynamics and show that a moonlet can open a gap by collisionless gravitational interaction alone through clearing of the chaotic zone in the moonlet vicinity \citep{Longaretti2}. In \citet{graetz2018} we derived an axisymmetric diffusion model, that describes the radial density profile of a gap an embedded moon creates. The gap profile is determined by an equilibrium between scattering of the ring particles (collisionless) and viscous diffusion back into the gap\footnote{Regarding the differences between the diffusion equations developed in \citet{sremvcevic2000density, sremvcevic2002density}, and \citet{graetz2018}: both models account for the two counteracting processes of gravitational scattering by an embedded perturber and viscous diffusion back into the created gap. \citet{sremvcevic2000density, sremvcevic2002density} describe propeller objects caused by moonlets that are too small to create a gap around the entire circumference of the ring. The scattering is modeled in the vicinity of the moonlet and the diffusion equation describes how the ring material diffuses back into the gap along the azimuth. \citet{graetz2018} describe radial density profiles of gaps caused by moons so heavy that they clear a complete gap. The timescales of diffusion and scattering can be separated, which permits the calculation of an azimuthally averaged surface density profile using an axisymmetric diffusion equation.}. In this article we will combine this description with the streamline formalism developed by \citeauthor{BGT82} to derive an axisymmetric diffusion equation that accounts for (1) the (collisionless) scattering of the ring particles by the embedded moon, (2) the viscous diffusion back into the gap, and (3) the angular momentum flux reversal that is responsible for the sharpness of the gap edges.\\

The article is organized as follows:
First, we briefly summarize the axisymmetric diffusion model we developed in \citet{graetz2018} to describe radial density profiles in the vicinity of embedded moons. Then we outline the streamline formalism introduced by \citet{BGT82} to describe the flux reversal. We use this streamline formalism to derive the pressure tensor of the ring and from that the shear averaged over one synodic period. Next, we insert this shear into the diffusion equation derived in \citet{graetz2018}, in order to account for the reversal of the angular momentum flux and we then finally apply the model to the Encke and Keeler gap to model their sharp gap edges (figure \ref{fig:uvis}) with the means of an axisymmetric nonlinear diffusion equation and to estimate the ring's shear viscosities in their vicinities.

\section{Model}
In \citet{graetz2018} we described the azimuthally averaged surface density, $\Sigma$, of a planetary ring in the vicinity of an embedded moon with a diffusion equation that accounts for the gravitational scattering  of the ring particles as they pass the moon and for the counteracting viscous diffusion that has the tendency to fill the created gap:
\begin{equation}
  \partial_t\Sigma+\partial_x\left(\Sigma \frac{\mathrm da}{\mathrm dt} -3\partial_x\nu\Sigma\right)=0\quad .\label{eq:non_lin_diff_eq_v2}
\end{equation}

Here, $a$ denotes the semi-major axis of the ring particles that pass the moon (located at $x=y=0$) and $\nu$ denotes the ring's shear viscosity.
For impact parameters larger than $5h$ the scattering of the passing ring particles can be described by $\mathrm da/\mathrm dt=\alpha\sign(x)/x^4$ \citep[equation 60]{gt82} with
\begin{equation}
\alpha=\mathscr A_1^2/(18\pi)\cdot \Omega \left(M_\text{s}/M_\text{p}\right)^2 a_0^5\label{eq:alpha}
\end{equation}
and $\mathscr A_1=6.7187$ \citep{gt82,Seiss2010}. $\Omega$ is the Kepler frequency and $M_\text{s}$ and $M_\text{p}$ denote the masses of the moon and the planet, respectively.
For smaller impact parameters, \citet{graetz2018} solved Hill's equations numerically, calculated the particles' scattering, and fitted the more general result
\begin{equation}
  \frac{\mathrm da}{\mathrm dt}_\text{fit}= \frac{ \alpha\sign(x)}{x^4+A\sign(x)hx^3+Bh^2x^2}\label{eq:dadt_fit}
\end{equation}
giving $A= 0.711557$ and $B = -7.58607$. Here, $h$ denotes the Hill radius:
\begin{equation}
  h=a_0\sqrt[3]{M_\text{s}/(3M_\text{p})}
\end{equation}

We described the viscosity of the ring with a power-law depending on the surface density, $\nu(x)=\nu_0\left(\Sigma(x)/\Sigma_0\right)^\beta$ \citep[equation 26]{schmit_tscharnuter_95}, where $\Sigma_0$ is the density of the undisturbed ring, and then solved equation \ref{eq:non_lin_diff_eq_v2} to find a radial density profile of:
\begin{equation}
  \Sigma(x)=\Sigma_0\left(\frac{\alpha\beta \sign(x) g\left(x\right)}{3\nu_0 (1+\beta)}+1\right)^{\frac 1\beta }\quad\text{with} \quad g(x)=\int\limits_{-\infty}^x\frac{1}{x^{\prime 4}+A\sign(x^\prime) h x^{\prime 3}+B h^2x^{\prime 2}} \,\mathrm  dx^\prime\label{eq:solution_powerlaw_simple}
\end{equation}

\citet{graetz2018} used this model to estimate the ring's shear viscosities in the vicinity of the Encke and Keeler gap (using the masses of Pan and Daphnis, their semi-major axes and the widths of the gaps) and to conclude that tiny single icy collision shards cannot be the cause for the numerous gaps that can be found in the C ring and the Cassini division: the creation of those rather wide gaps would require moonlet sizes, which could not have been overseen by Cassini -- even for viscosities as low as only $\SI{1}{\centi\meter\squared\per\second}$, which is lower than what has been measured in those regions using density waves \citep{Tiscareno200714,Colwell2009}.

In order to derive the diffusion equation \ref{eq:non_lin_diff_eq_v2}, we neglected scalar pressure and bulk viscosity and assumed that $\partial_x u_y\gg \partial_x u_x$ (where $u_x$ and $u_y$ are the radial and azimuthal velocities of the ring material) and could therefore approximate the pressure tensor by $\mathbf P_{xy}=\mathbf P_{yx}=3\nu\Sigma\Omega/2$. However, using these assumptions, our diffusion model cannot explain the formation of sharp gaps as they have already been observed by \emph{Voyager}, instead, the surface density gradually decreases to zero over several Hill radii.

In this article we calculate the pressure tensor using the streamline formalism proposed by \citet{BGT82} to account for the fact that the shear and thus the angular momentum transport is heavily modified in the wake region. We show that our streamline-diffusion model is able to explain that not only the local shear changes its sign in this wake region but that close to the moon even the angular momentum transport averaged over one synodic period (called angular momentum luminosity) reverses, which causes the normalized surface density of the ring to drop from order unity to zero over a range of only a few meters.

The plan of this section is as follows: First, we present the streamline formalism, then we calculate the radial derivatives of the particles' velocities that we need to derive the shear and pressure tensors. Using the pressure tensor, we calculate how the eccentricities that the moon induces as the particles pass by are damped using Gauss' perturbation equation. Knowing how the induced eccentricity behaves as a function of radial distance $x$ and time $t$ after the passing of the moon in turn allows us to calculate the shear as a function of $x$ and $t$. We then average the shear over one synodic period and insert this result into our diffusion equation. Finally, we show that a sharp gap edge forms at the radial distance from the moon where the averaged shear vanishes.
\subsection{Streamline formalism}
In general, the problem consists of an encounter of two particles orbiting around a central body (the general three-body problem). However, the problem can be severely simplified: The mass of the central body is much larger than the mass of the moon, which again, is much larger than the mass of the ring particle. Most of the time, both the moon and the ring particle move around the central body according to the laws of the two-body problem and only when they encounter each other is the ring particles' motion determined by both the planet's and the moon's gravitational field (the restricted three-body problem \citep{Spahn198769}). Furthermore, we assume a planar motion, so that the problem is reduced to two dimensions.

Under these circumstances the equations of motion of the ring particle can be linearized around the position of the moon (at $x=y=0$) which results in the so-called Hill equations:
\begin{align}
  \ddot{ x} -2\Omega_0\dot{y}- 3\Omega_0^2{x}&= F_x\label{eq:Hill_x}\\
  \ddot{  y} +2\Omega_0\dot{ x}&= F_y\label{eq:Hill_y}
\end{align}

Here, $\Omega_0$ denotes the Kepler frequency. The central gravity of the planet is expanded around the origin of the corotating coordinate system (the moons position), where the planet's gravity and the centrifugal force balance in the Hill equations. $F_{x,y}$ is the sum of the self-gravity of the ring and the gravitational force of the moon; however, here we neglect the self-gravity. Since the Hill sphere, which defines the region in which the moon’s gravitational influence dominates over that of the planet, is very small compared to the circumference of the disk \citet{Spahn1989124} argue that the model disk may be separated into two regions, a scattering region and the rest of the disk. Outside of the scattering region the gravitational force of the embedded moonlet onto the ring particle is always small compared to that of the planet. We therefore assume that an eccentricity is induced during close encounters with the moon and that outside of the scattering region $F_{x,y}\approx 0$. Outside of the scattering region the unperturbed solution of equations \ref{eq:Hill_x} and \ref{eq:Hill_y} is \citep{stewart91}:
\begin{align}
  x&=X-ea_0\cos\phi\\
  y&=Y+2ea_0\sin\phi\\
  \text{with}\quad \phi&=\Omega_0t-\tau\quad\text{and}\quad Y=Y_0-\frac 32X\Omega_0t.\label{eq:Y_and_phi}
\end{align}

Here, the coordinates in capital letters, $X$ and $Y$, give the time averaged position of the ring particle, denoted guiding center. The eccentricity, $e$, specifies the amplitude of the oscillation around the guiding center, $\phi$ denotes its phase \citep{stewart91}. Furthermore, $a_0$ is the semi-major axis and $\tau$ is the longitude of pericenter.

We assume that particles upstream of the moon have no initial eccentricity and can thus set $Y_0$ to $0$. As the particles pass the moon, an eccentricity, $e\sim X^{-2}$, is induced. Once outside the Hill sphere ($F_x=F_y\to 0$), the unperturbed particles perform an epicyclic motion around the guiding center that drifts with $-3X\Omega_0t/2$ in the azimuthal direction away from the moon. Such a trajectory, on which an ensemble of ring particles with the same impact parameter is assumed to move, is called a \emph{streamline}.

As ring particles with different impact parameters move on their respective streamlines, the shear causes the ring material to be compressed as the streamlines approach each other. The degree of radial compression of the ring material is given by \citep{SHOWALTER1986}:
\begin{align}
  J(\phi,X)&=\frac{\mathrm dx}{\mathrm dX}=1-a_0\frac{\mathrm de}{\mathrm dX}\cos\phi+ea_0\frac{\mathrm d\phi}{\mathrm dX}\sin\phi.\label{eq:compression_before_q}\\
  \intertext{By introducing a generalized nonlinearity parameter $q$ with phase $\gamma$}
  q\cos\gamma&\coloneqq -a_0\frac{\mathrm de}{\mathrm dX}\label{eq:cos_q}\\
  q\sin\gamma&\coloneqq -ea_0\frac{\mathrm d\phi}{\mathrm dX}\label{eq:sin_q}\\
  \intertext{equation \ref{eq:compression_before_q} becomes:}
  J&=1+q\cos\left(\phi+\gamma\right)\text{.}
\end{align}
The nonlinearity parameter $q$ (with $0\le q\le 1$) describes the streamline compression where $q=0$ means that the ring is undisturbed and $q=1$ that the streamlines intersect. Phase $\gamma$ describes the relative importance of the contributions of the derivative of the phase and the derivative of the eccentricity to the nonlinearity parameter $q$.

The changes in the surface density field of the ring, $\sigma(x,\phi)$, caused by the compression is given by \citep{BGT89,SPAHN1994}:
\begin{equation}
  \sigma(x,\phi)=\frac{\Sigma(x)}{J}.
\end{equation}

Let us estimate $\gamma$:
We identify $\mathrm d\phi/\mathrm d X$ with the radial wave number, $k=\mathrm d\phi/\mathrm d X$, and from equation \ref{eq:Y_and_phi} with $Y_0=0$ follows
\begin{align}
  \frac{\mathrm d\phi}{\mathrm d X} &=\frac{2Y}{3X^2}=-\frac{\Omega_0 t}{X}= k\text{.}\\
  \intertext{From $e\sim X^{-2}$ follows}
  \frac{\mathrm de}{\mathrm dX}&=-\frac {2e}X,\\
  \intertext{so that we can conclude using equations \ref{eq:cos_q} and \ref{eq:sin_q}:}
  \tan\gamma&=\frac{q\sin\gamma}{q\cos\gamma} = \frac{-ea_0\mathrm d\phi/\mathrm dX}{-a_0\mathrm de/\mathrm dX}=-\frac {Y}{3 X}.\label{eq:tan_gamma}
\end{align}

We assume a frame of reference centered at the embedded moon's position with an $x$-axis pointing outward and a $y$-axis pointing in the moon's orbit direction. In this frame of reference, the ring is disturbed for $x>0$ and $y<0$ (outer edge) and $x<0$ and $y>0$ (inner edge).

The particles drift away from the embedded moon toward $Y\to\pm\infty$, the right-hand side of equation \ref{eq:tan_gamma} tends to $+\infty$ and thus $\gamma\nearrow -\pi/2$ or $+\pi/2$. Since $q\ge 0$, the expression $q\sin\gamma=-ea_0 2Y/3X^2$ gives $\gamma\approx\pi/2$ for the disturbed region of the outer edge and $\gamma\approx-\pi/2$ for the disturbed region of the inner edge. The fact that $\gamma\approx\pm\pi/2$ expresses that, in the tight winding limit, in which the wake damping occurs, the derivative of the phase dominates over the derivative of the eccentricity regarding their contribution to the nonlinearity parameter $q$ \citep{Longaretti2}.

Next, we calculate the particles' velocities, $u$ and $v$, their radial derivatives, the shear tensor and then the pressure tensor. We perform those calculations for the outer gap edge, the solution for the inner gap edge can be obtained similarly but follows also from symmetry.

The velocities of the particles moving on the streamlines are:
\begin{align}
  u&=\frac{\mathrm dx}{\mathrm dt}=ea_0\Omega_0\sin\phi\\
  v&=\frac{\mathrm dy}{\mathrm dt}=-\frac 32\Omega_0X+2ea_0\Omega_0\cos\phi.
\end{align}

To calculate the shear tensor, we first have to determine the radial derivatives of the velocities:
\begin{align}
  \frac{\mathrm du}{\mathrm dx}&=\frac 1J\frac{\mathrm du}{\mathrm dX}=-\frac 1J\Omega_0q\sin\left(\phi+\gamma\right)\overset{\gamma\overset{!}{=}\frac\pi 2}{=}-\frac 1J\Omega_0q\cos\left(\phi\right)\\
  \frac{\mathrm dv}{\mathrm dx}&=\frac 1J\frac{\mathrm dv}{\mathrm dX}=\frac 1J\Omega_0\left(-\frac 32-2q\cos\left(\phi+\gamma\right)\right)\overset{\gamma\overset{!}{=}\frac\pi 2}{=}\frac 1J\Omega_0\left(-\frac 32+2q\sin\left(\phi\right)\right).
\end{align}

Now we can calculate the traceless shear tensor, neglecting azimuthal derivatives of the velocities, $u$ and $v$, due to the tight winding approximation ($\partial_r \bm u_i\gg \partial_\phi\bm u_i$):
\begin{align}
  \mathbf D&=\frac 12\left(\nabla\circ\bm u+\left(\nabla\circ\bm u\right)^T\right)-\frac 13\left(\nabla\cdot \bm u\right)\mathds 1\\
  \Rightarrow \mathbf D_{xx}&=-\frac 2{3}\frac 1J q \Omega_0\cos(\phi)\quad\text{and}\quad \mathbf D_{xy}=\frac 1{2}\frac 1J\Omega_0\left(-\frac 32+2q\sin\left(\phi\right)\right).
\end{align}

Here, $\mathds 1$ is the identity matrix. The expression for $\mathbf D_{xy}$ is equal to \citet[equation 129]{Lehmann_2016}.
Let us finally define the pressure tensor (Newton):
\begin{equation}
  \mathbf P=p\mathds 1-2\sigma\nu\mathbf D-\sigma\zeta\left(\nabla\cdot \bm u\right)\mathds 1.\label{eq:pressure_tensor}
\end{equation}

\citet{graetz2018} derived the azimuthally averaged radial density profile, $\Sigma(x)$ (equation \ref{eq:solution_powerlaw_simple}), assuming a viscosity of $\nu(x)=\nu_0\left(\Sigma(x)/\Sigma_0\right)^\beta$ (power law). Here, the surface density, $\sigma$, and the shear viscosity, $\nu$, are functions of radial distance from the moon, $x$, and azimuth, $\phi$, and we can write $\sigma(x,\phi)=\Sigma(x)/J$ because the surface density rises as the ring becomes compressed in the wake region. The viscosity therefore becomes $\nu(x,\phi)=\nu_0\left(\sigma(x,\phi)/\Sigma_0\right)^\beta$, and thus  $\nu(x,\phi)=\nu_0\left(\Sigma(x)/{\Sigma_0}\right)^\beta/{J^\beta}=\nu(x)/{J^\beta}$.

The scalar pressure, $p$, is proportional to the ring's surface density and the square of the sound speed, $c$, thus $p=\sigma(x,\phi)c^2=\Sigma(x) c^2/{J}$. We describe the bulk viscosity, $\zeta$, with a fixed ratio to the shear viscosity, $\nu$, with $\zeta=(\zeta_0/\nu_0)\nu$. With all this, the pressure tensor takes the following form \citep[Equation 137 (a), (b)]{Lehmann_2016}:

\begin{align}
  \mathbf P_{xx}&=\frac{\Sigma(x) c^2}{J}+\nu(x)\Sigma(x)\Omega_0\left(\frac 43 +\frac{\zeta_0}{\nu_0}\right)\frac 1{J^{2+\beta}}q\cos(\phi)\\
  \mathbf P_{xy}&=\nu(x)\Sigma(x)\Omega_0\frac 1{J^{2+\beta}}\left(\frac 32-2q\sin\left(\phi\right)\right)\text{.}
\end{align}

An explanation on how streamline crossing ($q = 1$ and thus $J=0$) is prevented in this model can be found below equation \ref{eq:f_different_betas}.

This pressure tensor can be modified for future treatment as follows:
First, we average the $xy$-component of the pressure tensor over one orbital period (short term), so that we can express the shear as a function of only the nonlinearity parameter, $q$. During this first averaging over the short time scale of one orbital period we assume $q$ and thus $e$ to be quasi-constant. This is a valid assumption because due to the short collisional time-scale, the stress tensor quickly reaches a quasi-equilibrium and then evolves quasi-statically with the large-scale dynamics \citep{Longaretti2}. Second, we calculate how the eccentricities (that are induced when the particles pass the moon) are damped, which allows us to calculate the nonlinearity parameter, $q(x,t)$, of the ring as a function of the radial distance to the moon, $x$, and the downstream time, $t$. Inserting $q(x,t)$ into $\langle\mathbf P_{xy}\rangle_\text{P}(q)$ then gives us $\langle\mathbf P_{xy}\rangle_\text{P}(x,t)$.

Finally, we average $\langle\mathbf P_{xy}\rangle_\text{P}(x,t)$ over the much longer time scale of one synodic period (where the temporal development of $e$, $q$ and thus the shear is considered) which gives us the averaged shear as a function of the radial distance from the moon, $\langle \mathbf P_{xy}\rangle_{T_\text{s}}(x)$, which is the desired result that we are going to insert into the diffusion equation.

\begin{figure}[h]
  \centering
  \figurenum{2}
  \epsscale{1}
  \includegraphics[width=18.75cm]{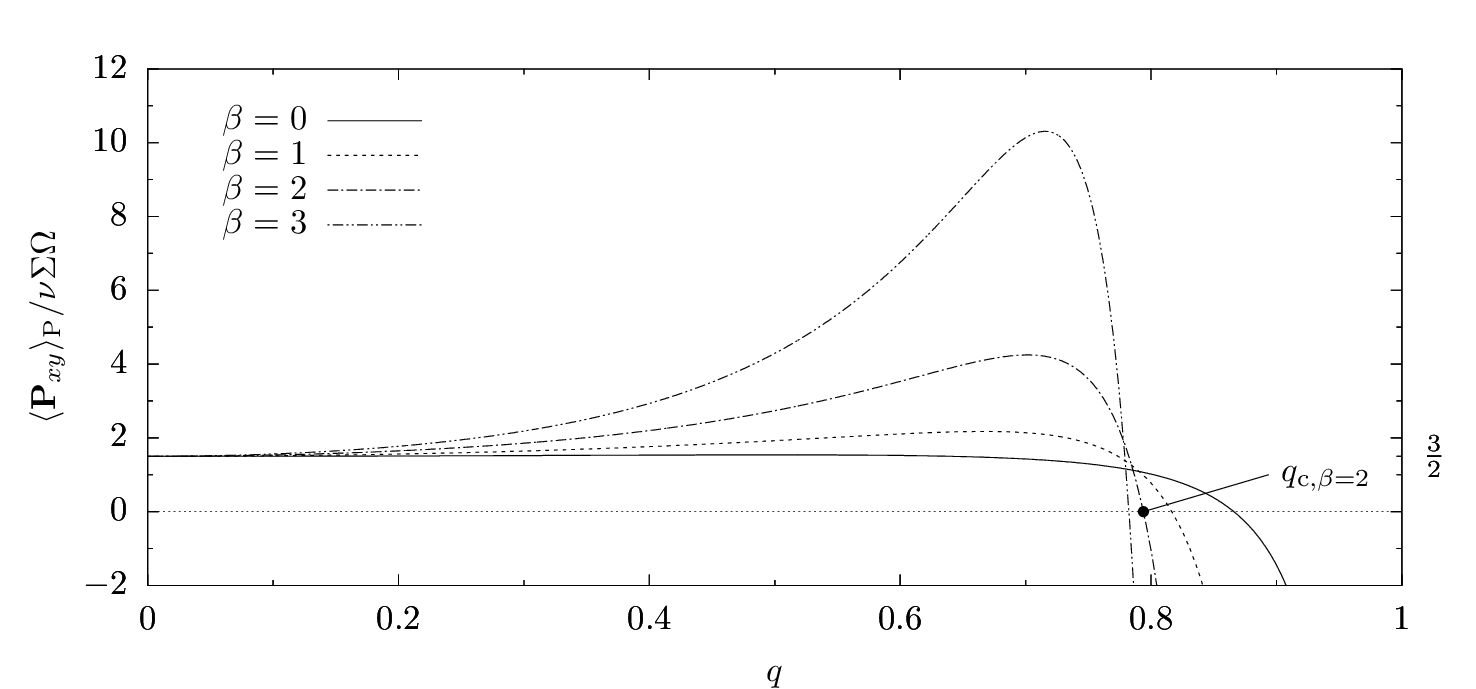}
  \caption{Orbit (short-term) averaged $xy$-component of the pressure tensor, normalized by $\nu(x)\Sigma(x)\Omega(x)$, as a function of the nonlinearity parameter, $q$, for different power-law exponents, $\beta$. For low $q$, the shear is Keplerian but increases and finally reverses for large values of $q$.}
  \label{pxy_mean_period}
\end{figure}

Let us start with averaging the $xy$-component of the pressure tensor over the short term
\begin{equation} 
  \langle \mathbf P_{xy}\rangle_\text{P}\left(q,\beta\right)=\frac 1{2\pi}\int\limits_{0}^{2\pi}\mathrm d\phi\, \mathbf P_{xy}\left(\phi,q,\beta\right)=\Sigma(x)\nu(x)\Omega_0
                                                        \begin{cases}
                                                          -\frac{4 q^2-3}{2 \left(1-q^2\right)^{3/2}}& \beta=0\\
                                                          \frac{6-9 q^2}{4 \left(1-q^2\right)^{5/2}}& \beta=1\\
                                                          -\frac{4 q^4+7 q^2-6}{4 \left(1-q^2\right)^{7/2}}&\beta=2\\
                                                          -\frac{51 q^4+8 q^2-24}{16 \left(1-q^2\right)^{9/2}}&\beta=3
                                                        \end{cases},
                                                        \label{eq:pressure_tensor_averaged_one_orbit}
                                                      \end{equation}
which can be written in algebraic form for integer values\footnote{The power-law exponent $\beta$ characterizes the transport of momentum in the ring. Nongravitating simulations by \citet{SALO1991} indicate a nearly linear $\nu$-$\Sigma$ dependence ($\beta\approx 1$) and \citet{schmit_tscharnuter_95} explain that $\beta=1.26$ is the ``realistic case'' in the sense that it corresponds to the shear viscosity coefficient calculated by \citet{ARAKI198683}. In such granular systems, the momentum transport is nonlocal due to finite size effects. \citet{Daisaka2001} showed that gravitational torques exerted by the collective motion of self-gravity wakes strongly enhance the viscosity indicating $\beta\approx 2$. In the presence of strong long-range correlations due to processes such as clumping in very dense regions, the momentum transfer might potentially be even more efficient yielding even higher values of $\beta$.} of $\beta$.
Figure \ref{pxy_mean_period} shows this averaged $xy$-component of the pressure tensor, $\langle \mathbf P_{xy}\rangle_\text P$, for different values of $\beta$. In the undisturbed regions of the ring, where $q\ll 1$, the shear is predominantly Keplerian with $\langle \mathbf P_{xy}\rangle_\text P=3/2\cdot\nu (x)\Sigma (x)\Omega_0$ but in the wake region downstream of the moon the streamlines come close to each other, $q$ rises, and $\langle \mathbf P_{xy}\rangle_\text P$ might even reverse at $q_c(\beta)$ if the ring is disturbed enough.

Next, we are going to calculate the damping of the induced eccentricities, $e$, downstream of the moon. The induced eccentricities are \citep{graetz2018}:

\begin{equation}
  \tilde e=\frac{\mathscr A_1}{\tilde X^2+m\left(\sign(\tilde X)\tilde X\right)^n}\quad\text{with}\quad \mathscr A_1=6.7187, m  = -54.8389, n  = -2.60934\label{eq:fit_e}
\end{equation}
with $\tilde X=X/h$ and they are damped according to the following orbit-averaged perturbation equation:
\begin{equation}
  \frac{\mathrm de}{\mathrm dt}=\frac 1{2\pi a_0 \Omega_0}\int\limits_0 ^{2\pi}\bigl(F_x \sin(\phi)+2F_y\cos(\phi)\bigr)\,\mathrm d\phi\text{.}
\end{equation}

The dissipative disturbing force is $\bm F=- \nabla\cdot \mathbf P/\sigma$ where derivatives in the azimuthal direction are neglected because the discussed structures are long stretched and tightly wound due to the Kepler shear, so that radial changes for density and mean velocities dominate over azimuthal changes (tight winding approximation). Remembering $\mathrm d\phi/\mathrm dX=k$, $\sigma=\Sigma/J$, and $J=\mathrm dx/\mathrm dX$ gives:
\begin{equation}
F_x=-\frac k{\Sigma(x)}  \frac{\mathrm d}{\mathrm d\phi}\mathbf P_{xx}\quad\text{and}\quad F_y=-\frac k{\Sigma(x)} \frac{\mathrm d}{\mathrm d\phi} \mathbf P_{xy}.
\end{equation}

Performing the orbit averaging of the perturbation equation using partial integration yields:
\begin{align}
  \frac{\mathrm de}{\mathrm dt}&=-\frac k{2\pi a_0\Omega_0\Sigma(x)}\int\limits_0^{2\pi}\left(\frac{\mathrm d\mathbf P_{xx}}{\mathrm d\phi}\sin(\phi)+2\frac{\mathrm d \mathbf P_{xy}}{\mathrm d\phi}\cos(\phi)\right)\,\mathrm d\phi\label{eq:int_t2_1}\\
                               &=\frac{k \nu(x)}{a_0}\frac {1}{2\pi}\int\limits_0^{2\pi}\left(\tilde{\mathbf P}_{xx}\cos(\phi)-2\tilde{\mathbf P}_{xy}\sin(\phi)\right)\,\mathrm d\phi\quad\text{with}\quad \mathbf P=\nu(x)\Sigma(x)\Omega_0\tilde{\mathbf P}\label{eq:int_t2_2}\\
                               &=\frac{k\nu(x)}{a_0}f(q,\beta,\frac{\zeta_0}{\nu_0})\overset{k=-\frac{\Omega_0 t}{X}}{=}-\frac{\nu(x)\Omega_0 t}{a_0 X}f(q,\beta,\frac{\zeta_0}{\nu_0}). \label{eq:flux_rev_e_num}
\end{align}

The integral in equation \ref{eq:int_t2_2} is equal to the viscous coefficient $t_2$ in \citet[equations 17, 18]{BGT_86}.
Since $q\sin\gamma=-e a_0 k$, $\gamma=\pi/2$, and $k=-\Omega_0 t/X$, we can calculate the nonlinearity parameter as a function of the eccentricity, $e$, downstream time, $t$, and impact parameter, $X$, with $q=ea_0\Omega_0 t/X$. We can neglect the expression $q\cos\gamma=-a_0\mathrm de/\mathrm dX$ because $\gamma=\pi/2$.
If we use scaled coordinates, $\tilde X=X/h$, $\tilde t=\Omega_0 t$, $\tilde e=a_0 e/h$, and $\tilde\nu(x)=\nu(x)/(h^2\Omega_0)$, we find:
\begin{align}
  \frac{\mathrm d\tilde e}{\mathrm d\tilde t}&=-\frac{\tilde t\tilde\nu(\tilde x)}{\tilde X}f\left(q,\beta,\frac{\zeta_0}{\nu_0}\right)\label{eq:ecc_over_t}\\
  q&=\frac{\tilde e\tilde t}{\tilde X}.\label{eq:q_from_e}
\end{align}

Please note that scalar pressure can be neglected when calculating $\mathrm de/\mathrm dt$, as $\int_0^{2\phi}\cos\phi/J(q,\phi)\,\mathrm d\phi=0$. For $\beta=0, 1, 2, 3$, the function $f\left(q,\beta,\zeta_0/\nu_0\right)$ has the following solutions:
\begin{align}
  f\left(q,\zeta_0/\nu_0\right)=\begin{cases}
    \frac{q^2 \left(3 \left(\sqrt{1-q^2}-1\right) \zeta_0/\nu_0-8 \sqrt{1-q^2}+11\right)-\left(\sqrt{1-q^2}-1\right) (3 \zeta_0/\nu_0-8)}{3 q \left(1-q^2\right)^{3/2}}& \beta=0\\
    \frac{q \left(q^2 (20-3 \zeta_0/\nu_0)+3 \zeta_0/\nu_0-11\right)}{6 \left(1-q^2\right)^{5/2}} & \beta=1\\
    -\frac{q \left(q^2 (3 \zeta_0/\nu_0-35)-3 \zeta_0/\nu_0+20\right)}{6 \left(1-q^2\right)^{7/2}} & \beta=2\\
     \frac{q \left(q^4 (44-3 \zeta_0/\nu_0)+q^2 (177-9 \zeta_0/\nu_0)+4 (3 \zeta_0/\nu_0-29)\right)}{24 \left(1-q^2\right)^{9/2}}& \beta=3\quad\text{.}\\
   \end{cases}
  \label{eq:f_different_betas}
\end{align}

Note that $f\left(q,\beta,\zeta_0/\nu_0\right)$ diverges at $q=1$, where the density becomes infinite in the streamline model, which leads to a drastic damping of the eccentricity (see equation \ref{eq:ecc_over_t}) causing $q$ to decrease (see equation \ref{eq:q_from_e}), thus preventing streamline crossing\footnote{\citet{stewart91} introduced the streamwire model, that was later used in \citet{Seiss2010} to avoid the singularity when the streamlines cross or merge.}.

With these results we are now able to calculate the short-term-averaged eccentricity, $e$, and nonlinearity parameter of the ring, $q$, as a function of the impact parameter, $X$, and downstream time, $t$. The nonlinearity parameter, $q(X,t)$, of the ring determines the shear, $\left\langle\mathbf P_{xy}\right\rangle_\text{P}(X,t)$, which we now finally average over one synodic period $T_\text{s}=4\pi a_0/(3\Omega_0 X)$ or $\tilde T_\text{s}=4\pi a_0/(3 \tilde X h)$:
\begin{align}
  \left\langle\mathbf P_{xy}\right\rangle_{T_\text{s}}(X)=&\frac 1{T_\text{s}(X)}\int\limits_0^{T_\text{s}(X)}\left\langle\mathbf P_{xy}\right\rangle_\text{P}(X,t)\,\mathrm dt\\
  \left\langle\mathbf P_{xy}\right\rangle_{T_\text{s}}(X)=&  \nu(X)\Sigma(X)\Omega_0\left\langle\tilde{\mathbf P}_{xy}\right\rangle_{T_\text{s}}(X)\\
\left\langle\tilde{\mathbf P}_{xy}\right\rangle_{T_\text{s}}(X)   \coloneqq& \frac 32 K(X).\label{eq:averaged_shear}
\end{align}

Since in the following we will be deriving an azimuthally averaged diffusion equation to describe a radial gap profile, we will approximate $K(X)=K(x)$.

\subsection{Hydrodynamic description}
In the next step we use this averaged shear (which is a function of $x$) to derive a diffusion equation that accounts not only for the gravitational scattering by the moon and the viscous diffusion of the ring, but also for the highly modified shear in the vicinity of the embedded moon that causes the angular momentum flux to reverse at the gap edge \citep{BGT82, BGT89}.

The derivation of the diffusion equation is similar to \citet{graetz2018}. Let us briefly summarize the steps: 
We start with the balances for mass and momentum that are given by the continuity and the Navier-Stokes equation using the thin-disk approximation with vertically averaged quantities:
\begin{align}
  \partial_t\Sigma+\nabla\cdot\left(\Sigma\bm u\right)&=0\\
  \Sigma\partial_t\bm u+\Sigma\left(\bm u\cdot\nabla\right)\bm u&=\Sigma\bm G-\nabla\cdot\mathbf P+\Sigma \bm f_m.
\end{align}

Here, $\bm G$, and $\bm f_{m}$ are the external accelerations and the acceleration exerted by the moon onto the ring material. In order to derive the diffusion equation, we reduce the azimuthal velocity to the leading Kepler term and assume the radial component to be much smaller, $\bm u=\left(u,-3/2\cdot\Omega x\right)$, where $x=r-a_0$ is the radial distance from the moon. The inertial accelerations are $\bm G=\left(2\Omega v+3\Omega^2x,-2\Omega u\right)$. We use equation \ref{eq:averaged_shear} For the $xy$-component of the pressure tensor.
In the following we are interested in radial structures ($\partial_y\to 0$): The azimuthal component of the Navier-Stokes equation can be rewritten in the form $\Sigma\partial_t v+\Sigma u\partial_xv=-2\Sigma\Omega u-\partial_x\mathbf P_{xy}+\Sigma f_{m,y}$ and thus with $v=-3/2\cdot\Omega x$ in the form:

\begin{equation}
  \Sigma u=\frac 2\Omega\left(\Sigma f_{m,y}-\partial_x\left(\nu(x)\Sigma(x)\Omega\frac 32 K(x)\right)\right).\label{eq:radial_mass_flux}
\end{equation}

The acceleration in the azimuthal direction exerted by the embedded moon on a ring particle with semimajor axis $a$ is
$\langle f_{m,y}\rangle=\Omega/2\cdot \mathrm d a/\mathrm dt$ \citep{graetz2018}. 
Inserting this radial mass flux \ref{eq:radial_mass_flux} into the continuity equation, we get a diffusion equation with an additional flux term accounting for the gravitational scattering of the ring material by the moon and for the ring's viscous diffusion, two counteracting physical processes that are able to create a gap. In comparison to Equation \ref{eq:non_lin_diff_eq_v2}, the additional term $K(x)$ in the following diffusion equation accounts for the flux reversal, which is responsible for the creation of sharp gap edges:

\begin{equation}
  \partial_t\Sigma+\partial_x\left(\Sigma\frac {\mathrm da}{\mathrm dt}-3\partial_x\Bigl(\nu(x)\Sigma(x)K(x)\Bigr)\right)=0.\label{eq:non_lin_diff_eq}
\end{equation}

In the limit of large impact parameters the scattering of the ring particles passing the moon is proportional to $x^{-4}$ \citep{gt80}. However, performing test particle simulations solving Hill's equations \citet{graetz2018} found that the scattering for small impact parameters is larger than what \citet{gt80} predict so that we decided to use expression \ref{eq:dadt_fit} that we obtained in \citet{graetz2018} by fitting the results of those test particle simulations.

In order to calculate the radial density profiles for the Encke and Keeler gap as shown in section \ref{sec:application} we solve the stationary case ($\partial_t\Sigma\to 0$) of equation \ref{eq:non_lin_diff_eq} numerically using a fourth-order Runge-Kutta method starting with $\Sigma/\Sigma_0=1$ at $x=50h$ for the Encke and $x=30h$ for the Keeler gap. The averaged shear $K(x)$ is calculated the following way: For each respective $x$ we calculate the eccentricity as a function of downstream time, $t$, solving equation \ref{eq:ecc_over_t} from $t=0$ to $t=T_\text{s}(x)$ with initial condition $e$ given by equation \ref{eq:fit_e}. From $e(x,t)$ we calculate the nonlinearity parameter of the ring, $q(x,t)$, using equation \ref{eq:q_from_e} and insert this result into equation \ref{eq:pressure_tensor_averaged_one_orbit} yielding $\left\langle \mathbf P_{xy}\right\rangle_\text{P}(x,t)$, which is finally averaged over one synodic period giving us the desired $K(x)$.

\section{Application}
\label{sec:application}
\begin{figure}[H]
  \centering
  \figurenum{3}
  \epsscale{1}
  \begin{minipage}{.43\linewidth}
    \includegraphics[width=.95\textwidth, trim={.5mm .5mm .5mm .5mm}, clip]{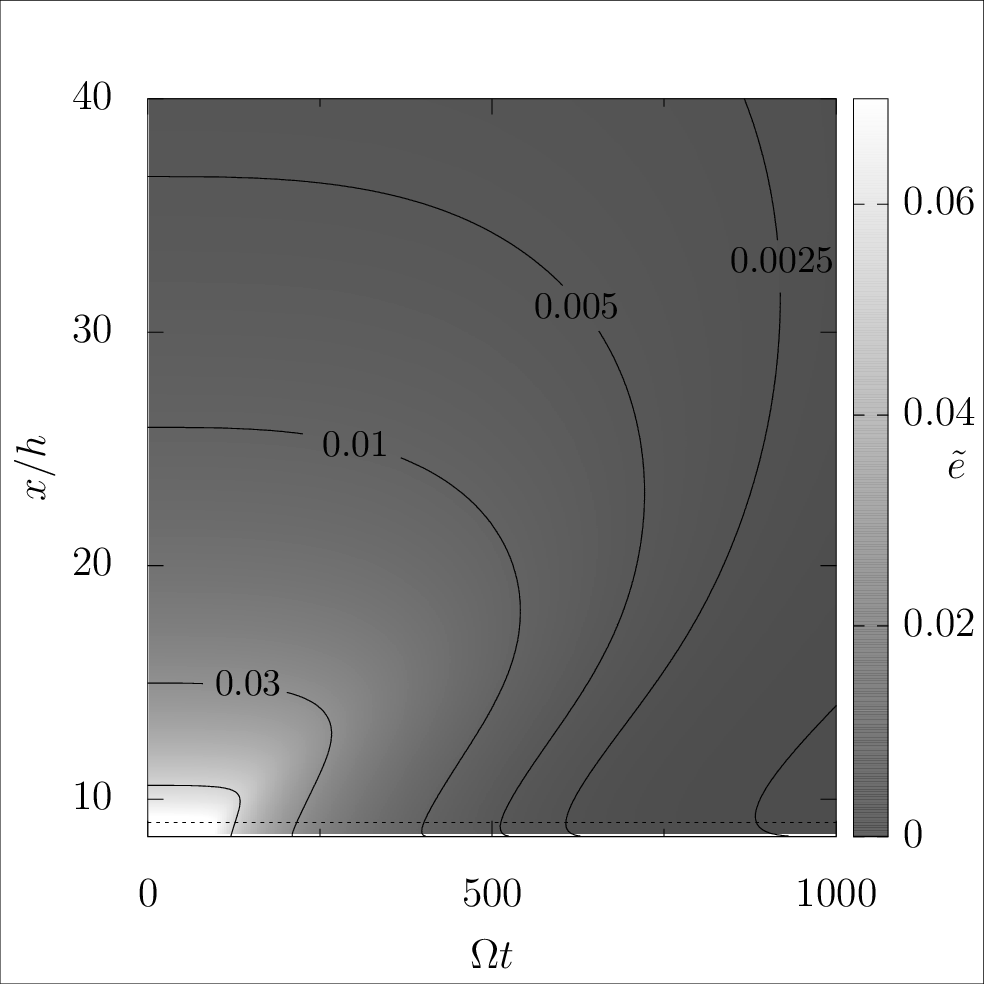}
  \end{minipage}
  \begin{minipage}{.43\linewidth}
    \includegraphics[width=.95\textwidth, trim={.5mm .5mm .5mm .5mm}, clip]{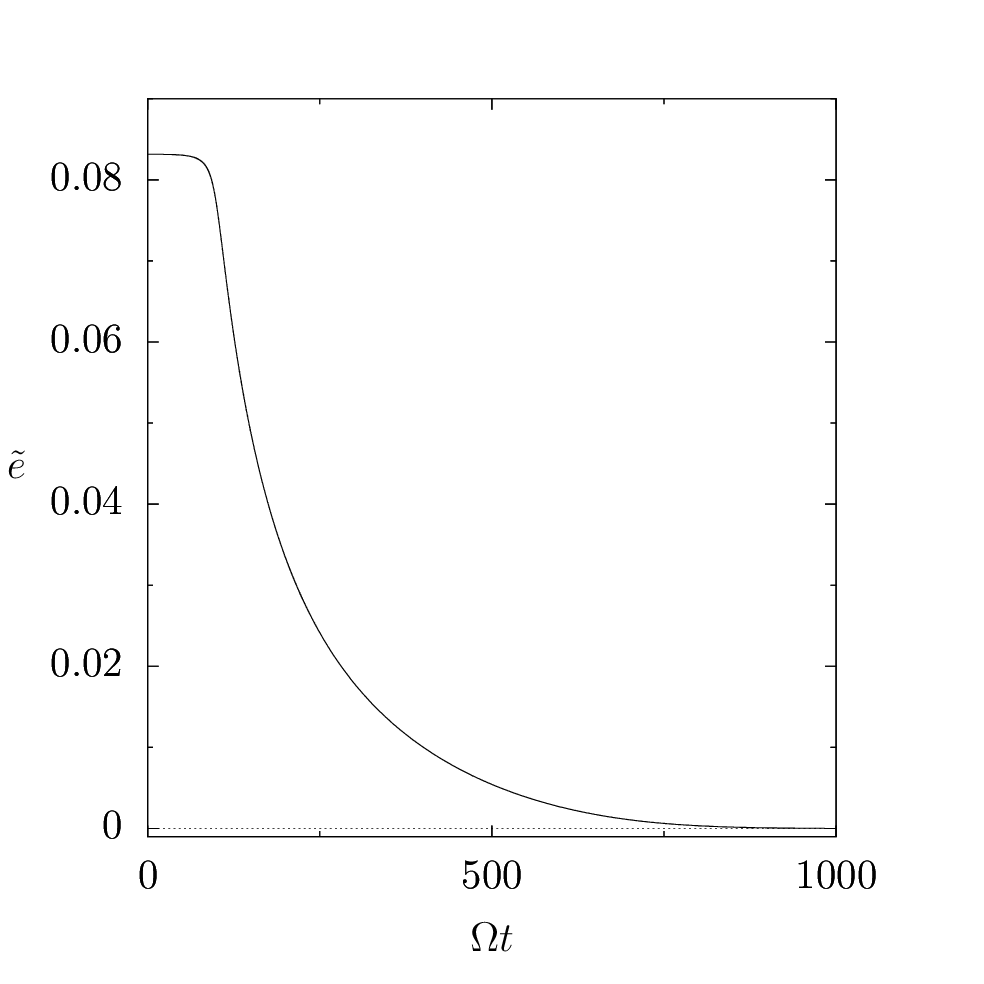}
  \end{minipage}\\
    \begin{minipage}{.43\linewidth}
    \includegraphics[width=.95\textwidth, trim={.5mm .5mm .5mm .5mm}, clip]{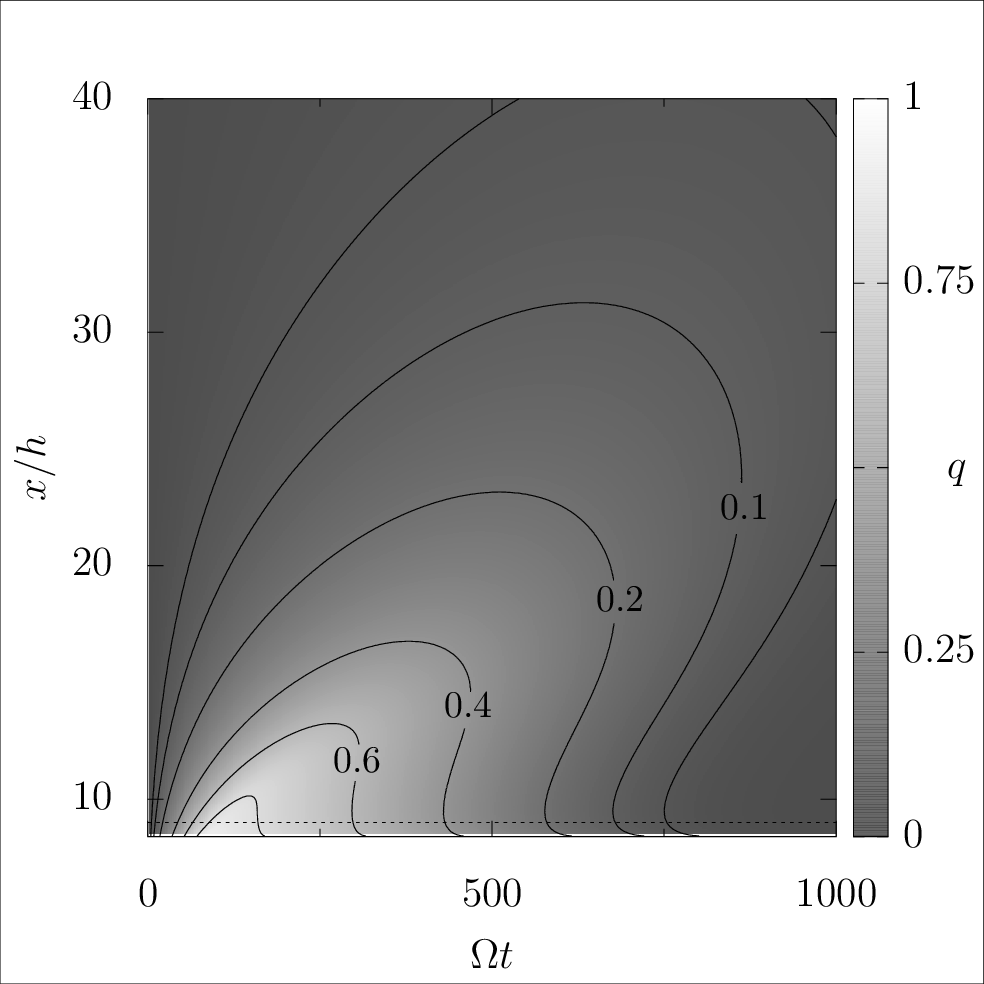}
  \end{minipage}
  \begin{minipage}{.43\linewidth}
    \includegraphics[width=.95\textwidth, trim={.5mm .5mm .5mm .5mm}, clip]{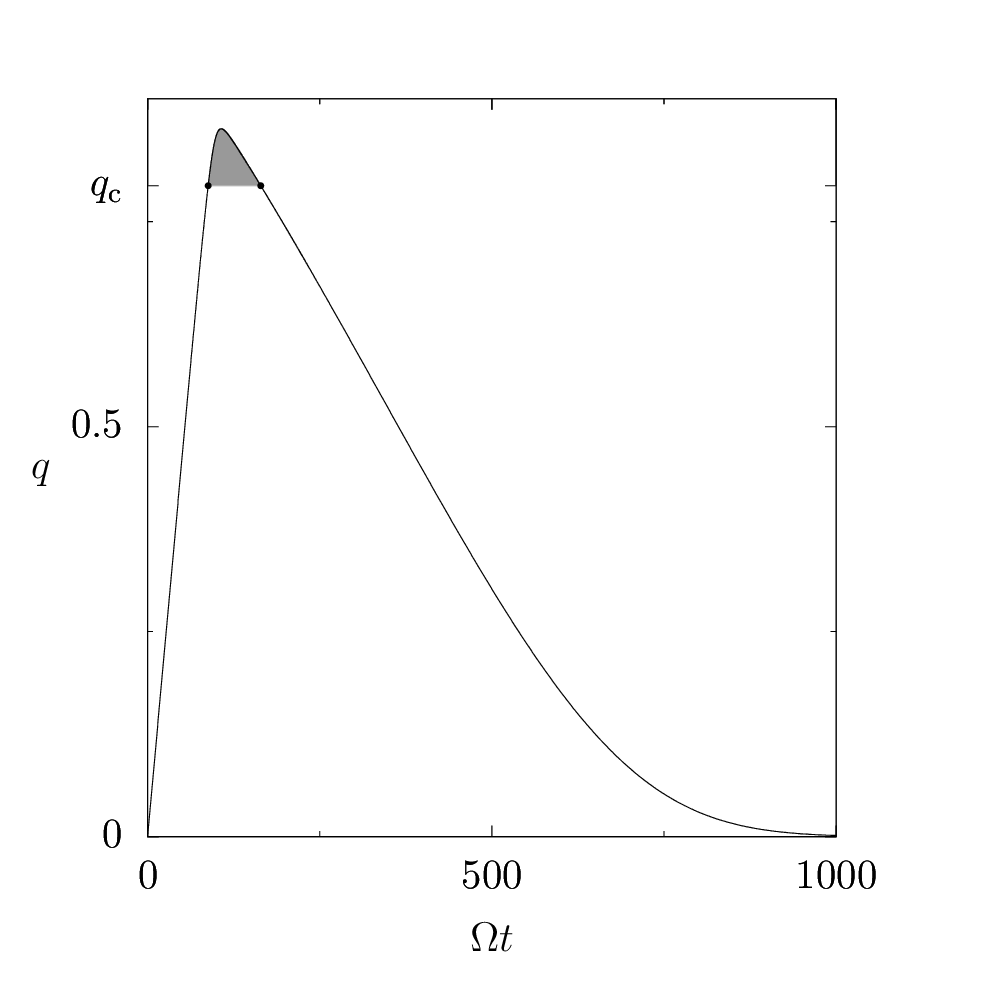}
  \end{minipage}\\
    \begin{minipage}{.43\linewidth}
    \includegraphics[width=.95\textwidth, trim={.5mm .5mm .5mm .5mm}, clip]{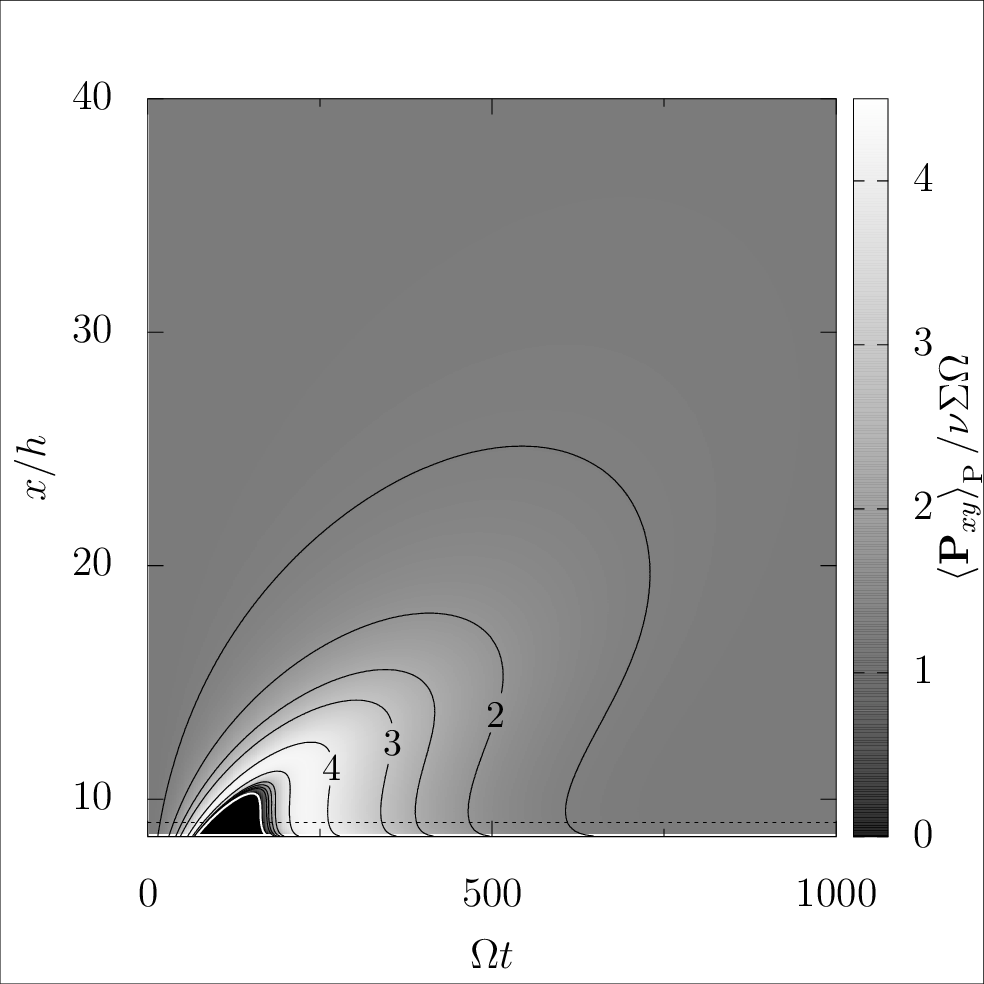}
  \end{minipage}
  \begin{minipage}{.43\linewidth}
    \includegraphics[width=.95\textwidth, trim={.5mm .5mm .5mm .5mm<}, clip]{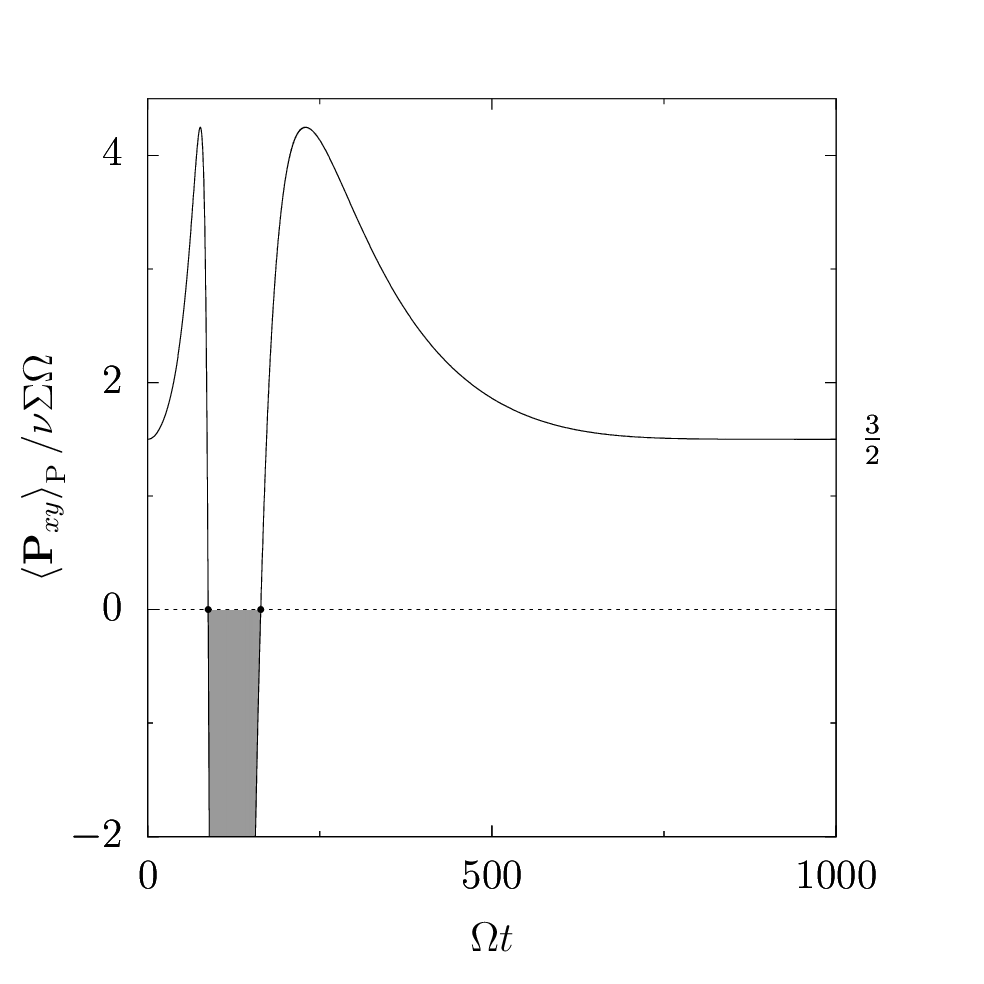}
  \end{minipage}
  \caption{The left column shows the eccentricity, $e$, the nonlinearity parameter, $q$, and the $xy$-component of the pressure tensor, $\bigl\langle\tilde{\mathbf P}_{xy}\bigr\rangle_\text{P}$, as a function of the radial distance of the moon, $x$, and the downstream time, $t$, after the ring particles passed the moon. The right column shows cuts through the plots in the respective left panels at $x=9h$. If $q$ rises above $q_\text{c}$, the shear reverses (region marked in gray in the right column). These plots correspond to the density profile of the Encke gap shown in the upper panel of figure \ref{fig:sharp-gaps} ($\beta=2$).}
  \label{fig:heat}
\end{figure}

As the ring material passes the moon (at $t=0$), an impulse in eccentricity $e$ is induced (see equation \ref{eq:fit_e}). For some time this induced eccentricity stays nearly constant; meanwhile the streamlines become compressed due to the shear and $q$ rises (see upper and middle right panels of figure \ref{fig:heat}). When the streamlines get to close to each other, the induced eccentricity of the ring particles is damped out through collisions, $e$ and $q$ decline\footnote{\emph{N}-particle simulations show that the damping of the eccentricity is not smooth but rather a step function with steep drops at the wake peaks and a fairly constant eccentricity between the peaks \citep[figure 6]{lewis_stewart_2000}. Our model is orbit-averaged and therefore shows a smooth decay of the induced eccentricity, which is in good agreement with the averaged eccentricity determined from \emph{N}-body simulations (compare the top right panel of figure \ref{fig:heat} to \citet{lewis_stewart_2000}'s Figure 5). }. However, if the streamline on which the ring particles are moving is close enough to the moon ($\tilde x\lessapprox 10$ in figure \ref{fig:heat}), something interesting happens: if we take a look at figure \ref{pxy_mean_period}, which shows $\bigl\langle\tilde{\mathbf P}_{xy}\bigr\rangle_\text{P}$ over $q$ we see that $\bigl\langle\tilde{\mathbf P}_{xy}\bigr\rangle_\text{P}$ first rises with $q$ then starts declining and reaches negative values above $q_c$. Taking a look at the plots of $q(t)$ and $\bigl\langle\tilde{\mathbf P}_{xy}\bigr\rangle_\text{P}(t)$ in figure \ref{fig:heat} (middle and bottom rows) we see that when the distortion of the streamlines is great enough and $q$ exceeds the corresponding critical value $q_c$ (region marked in gray), $\bigl\langle\tilde{\mathbf P}_{xy}\bigr\rangle_\text{P}$ becomes negative and the usually outward viscous flow of angular momentum is locally reversed, the material flows \emph{uphill} against the density gradient. In the bottom left panel in figure \ref{fig:heat}, which shows $\bigl\langle\tilde{\mathbf P}_{xy}\bigr\rangle_\text{P}$ as a function of $x$ and $t$, the white isoline denotes $\bigl\langle\tilde{\mathbf P}_{xy}\bigr\rangle_\text{P}=0$ and in the enclosed shark-fin-like black region, the $xy$-component of the pressure tensor has negative values. When averaging the shear over one orbital period, we find that it is greatly reduced in the vicinity of the moon and even reverses. This means that the ring is so strongly perturbed in the wake region, that not only the local shear but also the averaged angular moment transport over the entire synodic period is reversed. The gap edge constitutes itself exactly at the position where the averaged angular momentum flux changes its sign.

\begin{figure}[h]
  \centering
  \figurenum{4}
  \epsscale{1}
  \includegraphics[width=\textwidth]{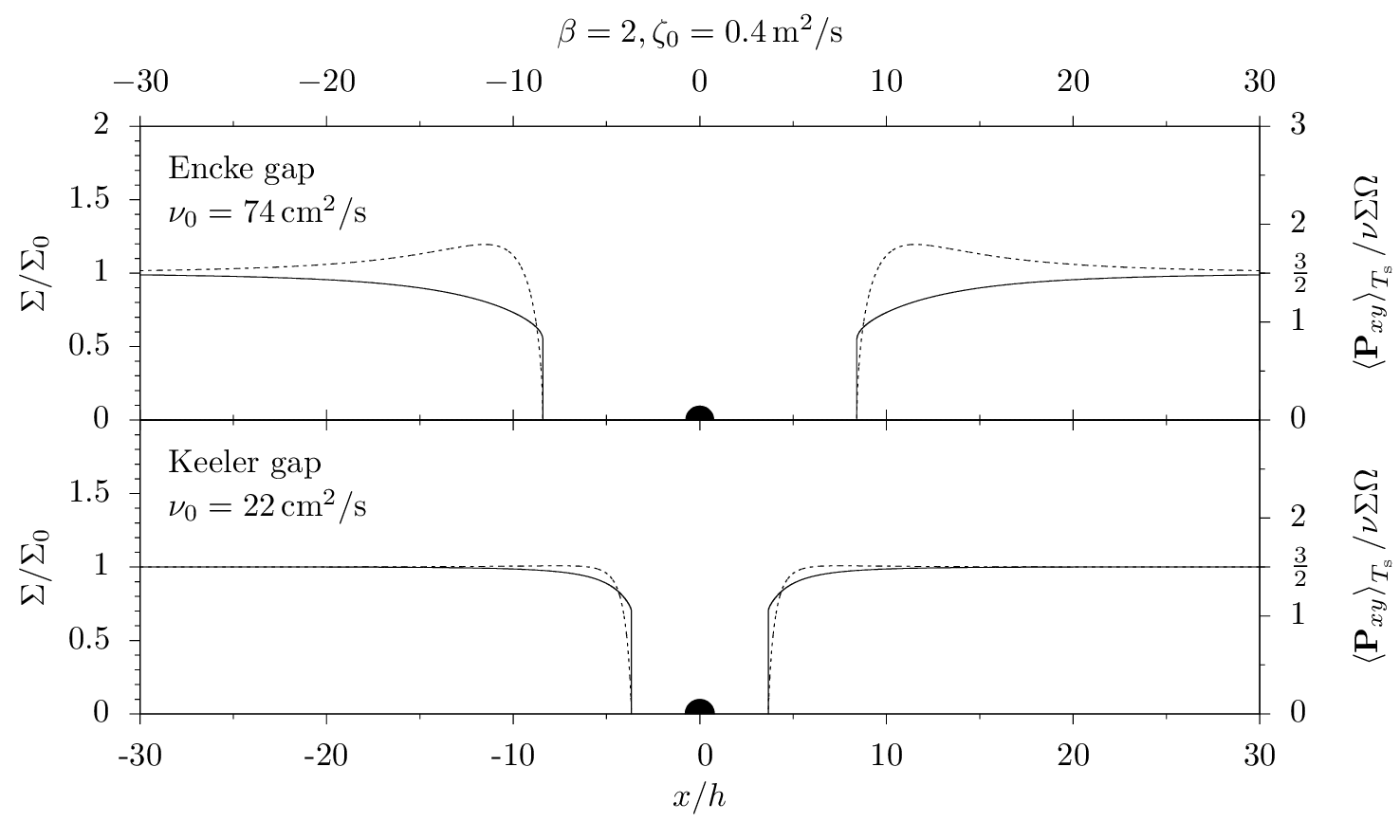}
  \includegraphics[width=\textwidth]{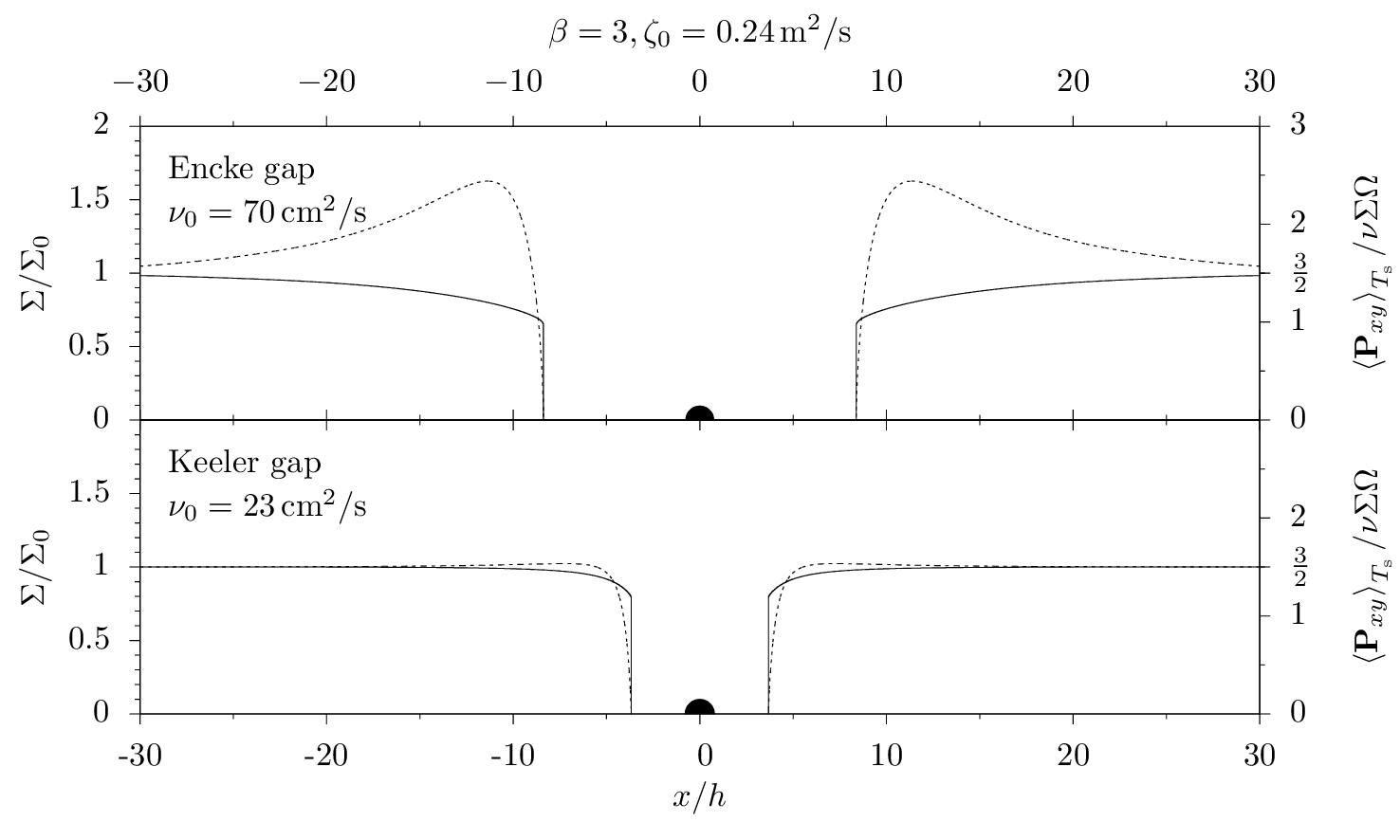}
  \caption{Radial density profiles (solutions of equation \ref{eq:non_lin_diff_eq}) for the Encke and Keeler gap for $\beta=2$ (as suggested by \citet{Daisaka2001}, upper plot) and $\beta=3$ (lower plot). The normalized surface density, $\Sigma(x)$, is depicted by a solid line, the $xy$-component of the pressure tensor averaged over one synodic period, $\bigl\langle\tilde{\mathbf P}_{xy}\bigr\rangle_{T_\text{s}}$, by the dashed lines. At the radial distance, where the shear vanishes, the diffusion of the ring is halted and a sharp gap edge forms.}
  \label{fig:sharp-gaps}
\end{figure}

Figure \ref{fig:sharp-gaps} shows radial density profiles for Encke and Keeler gap conditions as solutions of equation \ref{eq:non_lin_diff_eq}. Far away from the embedded moon the normalized surface density $\Sigma(x)$ is $\Sigma_0$ and the shear is undisturbed. Getting closer to the moon, the averaged shear first increases slightly and then drops to zero because the local shear is greatly disturbed in the wake region (see bottom row of figure \ref{fig:heat}). At the radial distance where the averaged angular momentum flux vanishes, the diffusion of the ring that tends to fill the gap is halted and the normalized surface density instantly drops from order unity to zero\footnote{The description of gap edges using a scalar viscosity (depending on the surface density via a power law) is a pragmatic simplification \citep{Spahn2000657, 2001Icar..153..295S} but the transport coefficients should in fact rather have a tensorial character due to a strong anisotropy at the gap edges. While the transport of material in the radial direction is halted where the angular momentum flux reverses, transport perpendicular to the ring plane is still possible, i.e. vertical splashing as observed in \emph{N}-body simulations \citep{lewis_stewart_2000}.} \citep{BGT82}.

The parameters for figure \ref{fig:sharp-gaps} are set as follows: The width of the gap is determined by the embedded moon's scattering and the counteracting viscous diffusion back into the gap. First, parameter $\alpha$ (see equation \ref{eq:alpha}), which describes the strength of the scattering by the moon (see equation \ref{eq:dadt_fit}) is calculated from the perturber's and the planet's masses and the perturber's semi-major axis. Then, the shear viscosity $\nu_0$ is adjusted, so that the resulting gap profile obtained by solving equation \ref{eq:non_lin_diff_eq} has the observed width of the respective gap.

For the masses $M_\text{Pan}=\SI[parse-numbers=false]{(4.95\pm 0.75)\cdot 10^{15}}{\kilo\gram}$ and $M_\text{Daphnis}=\SI[parse-numbers=false]{(8.4\pm 1.2)\cdot 10^{13}}{\kilo\gram}$ \citep[Table 1]{Porco1602} and $\beta=2$ we find $\SI[parse-numbers=false]{74\pm 22}{\centi\meter\squared\per\second}$ for the Encke gap and $\SI[parse-numbers=false]{22\pm6}{\centi\meter\squared\per\second}$ for the Keeler gap. For $\beta=3$ we find $\SI[parse-numbers=false]{70\pm22}{\centi\meter\squared\per\second}$ for the Encke gap and $\SI[parse-numbers=false]{23\pm 7}{\centi\meter\squared\per\second}$ for the Keeler gap. The error bars are a result of the uncertainties of the masses of Pan and Daphnis.

If one reduces the shear viscosity from $74$ to $\SI{50}{\centi\meter\squared\per\second}$ (all other parameters being equal to those used for the modeling of the Encke gap in the upper panel of figure \ref{fig:sharp-gaps}), the width of the modeled Encke gap increases from $320$ to $\SI{355}{\kilo\meter}$. If instead one increases the shear viscosity to $\SI{100}{\centi\meter\squared\per\second}$, the width of the modeled Encke gap reduces to $\SI{293}{\kilo\meter}$.
  
Our nonlinear wake damping model prevents streamline crossing ($q\ge 1$). However, for a too small value of the bulk viscosity we find that the perturbations are not damped quickly enough (the nonlinearity parameter $q$ exceeds the critical value $q_c$ for too long and gets too close to $1$). This has the effect that the shear reverses too far away from the moon, the moon is thus not able to push the material out of the gap and away from the point of flux reversal and the surface mass density starts to diverge. In this vain, we fixed the values for the bulk viscosity $\zeta_0$ for figure \ref{fig:sharp-gaps} to the minimum value that allows confinement of both the Encke and Keeler gap edges (the latter one requiring larger values). The chosen bulk viscosities are thus minimum values for the model to work, higher values would work as well.

The radial gap profiles for $\beta=2$ in the upper panel in figure \ref{fig:sharp-gaps} have been obtained for $\zeta_0=\SI{0.4}{\meter\squared\per\second}$ and therefore ${\zeta_0}/{\nu_0}\approx 54$ for the Encke and ${\zeta_0}/{\nu_0}\approx 180$ for the Keeler gap and the gap profiles for $\beta=3$ in the lower panel for $\zeta_0=\SI{0.24}{\meter\squared\per\second}$, thus ${\zeta_0}/{\nu_0}\approx 34$ and ${\zeta_0}/{\nu_0}\approx 104$ respectively. We note, however, that in a model with a more realistic equation of state it would be a diverging pressure due to dense packing that prevents streamline merging (see discussion in section \ref{sec:discussion}). \emph{N}-body simulations using hard inelastic spheres \citep{2001Icar..153..295S} suggest ${\zeta_0}/{\nu_0}$ around $4$.

\clearpage
\section{Discussion}
\label{sec:discussion}
UVIS-scans of the gap edges of the Encke and Keeler gap we presented (figure \ref{fig:uvis}) show that the photon count increases over a range of only around $\SI{100}{\meter}$ to the photon count of the unocculted star. The incorporation of the angular momentum flux reversal in the vicinity of the moon into the diffusion model for gaps that we developed in \citet{graetz2018} allows the description of such sharp gaps with the means of a viscous shear-diffusion equation.

The shear viscosity $\nu_0$ predominantly determines the position of the edge relative to the moon: without considering the flux reversal, we found that shear viscosities of $\SI[parse-numbers=false]{78\pm 24}{\centi\meter\squared\per\second}$ and $\SI[parse-numbers=false]{20\pm 6}{\centi\meter\squared\per\second}$ are needed to form the Encke and Keeler gap ($\beta=2$, see \citet{Daisaka2001}). If we consider the modified shear we find shear viscosities of $\SI[parse-numbers=false]{74\pm 22}{\centi\meter\squared\per\second}$ and $\SI[parse-numbers=false]{22\pm6}{\centi\meter\squared\per\second}$ for $\beta=2$, thus confirming the viscosities we estimated using the simplified model presented in \citet{graetz2018}. The error bars are a result of the uncertainties of the masses of Pan and Daphnis \citep[Table 1]{Porco1602}. The diffusion equation we derived in this article (Equation (\ref{eq:non_lin_diff_eq})) is more general than \citet[Equation (29)]{graetz2018} as it contains the additional expression $K(x)$ that accounts for the reversal of the thermodynamic force in the vicinity of the embedded perturber, which allows the description of much more strongly perturbed regions in dense planetary rings or protoplanetary disks.

The power law we used to describe the shear viscosity's dependency on the surface mass density is a parameterization introduced by \citet{schmit_tscharnuter_95} for small intervals of surface mass density. It is a reasonable simplification often used in hydrodynamic descriptions of planetary rings \citep{seiss2011}. However, it might prove to be very interesting to further analyze the viscosity's dependency on the surface mass density with a kinetic approach, especially in the limit of small, vanishing densities that occur at gap edges.

The ratios of bulk to shear viscosity we determine from our model are much larger than the one determined from \emph{N}-body simulations using hard inelastic spheres \citep{2001Icar..153..295S}. In our model, a higher bulk viscosity has the effect that the induced eccentricities get damped more quickly because more energy is dissipated in the wakes and the nonlinearity parameter of the ring $q$ exceeds the critical value $q_\text{c}$ at which the local shear reverses for a shorter period of time. If low values for $\zeta_0$ are chosen in our model, the flux reverses farther away from the moon and the surface density (instead of dropping to zero where the flux reverses) starts to diverge. In order to calculate the radial gap profiles in figure \ref{fig:sharp-gaps}, the bulk viscosity $\zeta_0$ has been chosen high enough to model the confinement of both the sharp Encke and the sharp Keeler gap edges, the latter requiring higher values.
On the one hand, one would indeed expect enhanced values of the bulk viscosity in a dense planetary ring when compared to the simulations that neglect particle rotation and coagulation of ring particles. Namely, a time-lag between excitations of the translational random motion of the particulate system and the spin temperatures typically contributes to the bulk viscosity \citep{ChapmanCowling1970book}\footnote{\citet{ratios} report that even common diatomic gases are seen to have bulk viscosities that are hundreds or thousands of times larger than their shear viscosities (if the molecules have vibrational or rotational modes that relax slowly).}.
Moreover, in the processes of coagulation and fragmentation, that are probably steadily occurring in dense rings \citep{spahn2004, Brilliantov9536}, the contact number of ring particles is constantly changing, which irreversibly alters the particles' configuration. This amounts to an additional process that dissipates random walk kinetic energy, in this way further increasing the value of the bulk viscosity.

On the other hand, we warn against interpreting the values we obtained for $\zeta_0$ literally, because our simple hydrodynamic description does not account for the fact that (especially in the wake region, where the streamlines converge) the ring cannot be compressed more than to a dense packing. Since in the pressure tensor (equation \ref{eq:pressure_tensor}) the bulk viscosity couples to the compression $\nabla\cdot\bm u$, it might be the case that the large bulk viscosities we find effectively enforce solutions with $\nabla\cdot\bm u \approx 0$ which would mean that the model for the outer A ring behaves incompressibly. Physically, in the rings this behavior might rather be established by an equation of state $\propto (\sigma-\sigma_\text{crit})^{-1}$ that yields a diverging pressure when the granular ring matter approaches a critical density $\sigma_\text{crit}$. \citet{BGT85} developed a granular flow model for dense planetary rings in which they introduced hydrodynamic transport coefficients. They modeled the rings as incompressible fluids in which the particles are densely packed. Since the model is three-dimensional, it allows splashing into the vertical direction when the material is compressed. The resulting dynamics of the local ring height can play an important role in the development of the moon induced wakes and the damping of large nonlinearity parameters $q$, which is not taken into account in our model. \citet{lewis_stewart_2000} observed evidence for vertical splashing of particles out of the ring plane at the wake peaks in detailed \emph{N}-particle simulations. They explain that this vertical splashing violates assumptions made in many analytic treatments of this region (including this one) but does not appear to invalidate the main conclusions drawn from such models.

In principle it is worth noting that the simple linear Onsager ansatz to quantify the angular
momentum flux does not sufficiently describe the situation of dense sheared granular media, and
probably nonlinear relations between shear and angular momentum flux should be tested in future studies.

The diffusion model we presented in \citet{graetz2018} and in this article allows the description of sharp gaps in planetary rings but can also be applied to other cosmic disks on completely different size scales such as protoplanetary disks. The model explains the sharp gap edges very well. From the width of the gap and the mass and semi-major axis of the embedded body the shear viscosity of the ring can be estimated. The high values of the bulk viscosity needed in this model make sense as it is generally accepted that the ring particles are aggregates that allow for the dissipation of large amounts of random walk kinetic energy when colliding. However, further research is needed to better understand the parameter bulk viscosity in hydrodynamic descriptions of cosmic disks.

\section{Acknowledgments}
This work was supported by Studienstiftung des deutschen Volkes and by Deutsches Zentrum f\"ur Luft- und Raumfahrt (OH 1401).


\bibliography{literatur_diss}



\end{document}